\documentstyle[aps]{revtex}
\begin{document} 
\draft
\title{ Universality of monopole mode and time evolution of a $d$-dimensional trapped 
interacting Bose gas }
\author{ Tarun Kanti Ghosh } 
\address {The Institute of Mathematical Sciences, C. I. T. Campus, Chennai-600 113, 
India.} 
\date{\today}
 
\maketitle
\begin{abstract}

We study a generalised Gross-Pitaevskii equation describing a 
$ d $-dimensional harmonic trapped (with trap frequency 
$\omega_{0}$) weakly interacting Bose gas with a
non-linearity of order (2 $ k $ + 1) and scaling exponent 
($ n $) of the interaction potential. Using the time-dependent 
variational analysis, we explicitly show that for a particular 
combination of $ n $, $ k $ and $ d $, the generalised 
GP equation has the universal monopole oscillation frequency $ 2 \omega_{0} $.
We also find that the time-evolution of the width can be 
described universally by the same Hill's equation if the system
satisfy that particular combination. We also obtain the condition for the 
exact self-similar solutions of the Gross-Pitaevskii equation.
As an application, we discuss low dimensional trapped Bose 
condensate state and Calogero model.

\end{abstract}

\noindent

\section{introduction}
There has been renewed interest in the Bose-Einstein condensation (BEC) after its 
experimental achievement by several groups \cite{anderson}.
The Gross-Pitaevskii (GP) equation explains most of the 
properties of these trapped dilute interacting Bose particles \cite{dal}. 
After the discovery of BEC in a trapped alkali atom, the influences of the 
dimension of a Bose system is being the subject of extensive study 
\cite{pit,ma,pet,ghosh,tarun}. Recently, low dimensional BEC has been observed
in MIT \cite{gus}.
It is well known that the nonlinear GP equation for a 2D trapped Bose system has a 
monopole mode with a universal frequency $ 2 \omega_0 $ \cite{pit,pitae,kagan}.
In 2D trapped interacting Bose system, the Fermi pseudopotential interaction does 
not affect the monopole mode frequency. This peculiar behavior of monopole 
mode frequency depends on the scaling property of the interaction potential and 
the order of non-linearity.
Without trap potential, the GP equation is scale invariant. Introduction of 
harmonic potential breaks the scale invariance. Pitaevskii and Rosch 
\cite{pit} has pointed out that 
due to special property of the harmonic potential, SO(2,1) symmetry still exists. 
They have shown the universal nature of the monopole mode frequency
only for 2D GP equation by constructing the generators of the SO(2,1) symmetry \cite{pit}.
Here, using the time-dependent variational approach, we obtain a 
condition for the universality
of monopole mode frequency and the dynamics of width of a class of GP equation
describing the trapped interacting Bose system, at varying spatial
dimensionality, order of the nonlinearity and the scaling
exponent of the interaction potential. Interestingly, the dynamics of the width 
of a class of GP equation can be described by the same Hill's equation if the
Bose system satisfy that particular combination or shows the universal monopole
mode frequency.

This paper is organised as follows. In Sec. II, using the time-dependent variational
analysis, 
we show that for a particular combination of $ n $, $ k $ and $ d $ the generalised GP equation
describing the $ d $-dimensional harmonic trapped Bose gas
exhibits the universal monopole oscillation frequency 2 $ \omega_0 $. We also
derive the time-evolution of the width of the $ d $-dimensional trapped Bose system 
which satisfies that particular combination. The time evolution of the width can be 
described universally by the same Hill's equation which is analytically solvable.
We also discuss the condition for exact self-similar solutions of the GP equation.
In Sec.III, we consider quasi-2D trapped Bose gas interacting through
the Fermi pseudopotential which can be described by an {\em ordinary } GP equation. 
In Sec. IV, we also consider the 1D {\em modified } GP equation which 
describes the BEC state. 
We obtain the universal monopole mode frequency and  time evolution of 
the condensate width of this system in a time-independent trap as well as a 
time-dependent trap. In Sec. V, we discuss the Calogero type potential. 
In Sec. VI, we present the summary and conclusions of our work.

\section{ time-dependent variational analysis}

To show the universality of the monopole mode frequency and the dynamics 
of the width of a class of GP equation (also known as nonlinear Schrodinger 
equation (NLSE)), we use the time dependent variational method.
This variational method has been extensively and successfully used
to explain many properties in BEC and other nonlinear systems.
Using this method, Garcia {\em et al}. \cite{gar} derived a general dispersion 
relation for the low energy excitations of a 3D deformed trapped Bose gas 
which exactly coincides with the results obtained by Stringari \cite{strin}
for large number of particles.

 Consider a system of Bose particles in $d$-dimensional harmonic trapped potential
 $ V_{t}( \vec r ) = \frac{1}{2} M \omega_{0}^2 r^2 $  at zero temperature and 
interacting with a translationally invariant interaction potential 
$ V_{I}= g_{d} V(\vec r -  \vec r^{\prime} ) $, where $ r^2 = \sum_{i=1}^{d} x_{i}^2 $
and $g_{d}$ is the coupling constant in $ d $ dimension. M is the atomic mass and
$ \vec r $ is a $ d $-dimensional vector.
Under the scale transformation, $ \vec r \rightarrow \lambda \vec r $, we assume
$ V_{I} (\lambda \vec r) = \lambda^n V_{I} (\vec r) $ where $ n $ is the scaling 
exponent of the  interaction potential.

We are considering a generalized  $d$-dimensional GP equation \cite{gross} 
describing a $ d $ -dimensional harmonic trapped weakly interacting Bose gas 
with a non-linearity 
of order $ (2k + 1 ) $  and scaling exponent $ n $ of an arbitrary interaction
potential $ V_I = g_d V (\vec r - \vec r^{\prime}) $
 
\begin{equation}\label{sch}
i \hbar \frac{\partial \psi ( \vec r )}{\partial t}   =  [ - \frac{ \hbar^2 }{2M} 
\nabla^2 + V_{t}(r) + g_d \int d^m r^{ \prime }
 \psi^*(\vec r^{\prime}) V (\vec r - \vec r^{\prime}) 
\psi ( \vec r^{\prime }) | \psi (\vec r )|^{2(k-1)}] \psi (\vec r), 
\end{equation}
where $ \nabla^2 $ is the $ d $-dimensional Laplacian operator and  
* denotes  complex conjugation.
The generalised $d$-dimensional energy functional is

\begin{equation}
 E[\psi]  =   \int d^d r [\frac{\hbar ^2}{2M} | \nabla \psi |^2 + V_{t} | \psi |^2 +
 \frac{g_d}{k+1} \int d^m r^{ \prime } \psi^*(\vec r) \psi^* (\vec r^{\prime })
V ( \vec r - \vec r^{\prime }) \psi (\vec r^{\prime }) | \psi (\vec r )|^{2(k-1)} \psi (\vec r)].
\end{equation}

If $ m = d $ = arbitrary dimension  and $ V(\vec r -\vec r^{\prime})$ is a 
$d$-dimensional delta function potential, the above equation gives 
the non-linear term of order $ (2k+2)$.
If $ k=1 $  and $ m=d \ge 1 $, Eq. (\ref{sch}) becomes {\em ordinary } GP equation.
and if $ k = 2 $ and $ m = d = 1 $, Eq. (\ref{sch})  gives the 1D {\em modified } 
 GP equation in which the $ |\psi |^6 $ interaction
term is present \cite{rybin,kolo}.

One can write down the generalised Lagrangian density corresponding 
to this system which is

\begin{equation}\label{lag}
L   =  \frac{i \hbar}{2} (\psi \frac{\partial \psi^*}{\partial t} -
\psi^* \frac{\partial \psi}{\partial t}) + [\frac{ \hbar^2}{2M} | \nabla \psi |^2 + 
 V_t (r) | \psi (\vec r ) |^2 + \frac{g_d}{(k+1)}\int d^m r^{\prime } 
 \psi^*(\vec r ) \psi^*(\vec r^{\prime })
V(\vec r - \vec r^{\prime}) \psi (\vec r^{\prime}) | \psi (\vec r )|^{2(k-1)} \psi (\vec r)].
\end{equation}

Without interaction potential, Eq. (\ref{sch}) reduces to the linear Schrodinger equation which has 
the Gaussian wave function for the ground state. In order to get the evolution of the square of the 
radius, we assume the following Gaussian wave function,
\begin{equation}\label{wave}
\psi (r) = C (t)  e^{- \frac{r^2}{2 a_{0}^2 }[\frac{1}{\alpha (t)} + i \beta (t)]},
\end{equation}
where $ C (t) $ is the normalization constant which can be determined by the 
normalization condition $ \int | \psi |^2 d^d r = N $  and 
 $ a_0 = \sqrt {\hbar / M \omega_0 } $ is the oscillator length. 
N is the total number of particles in the system. 
$ \alpha (t) $ and $ \beta(t) $ are the time-dependent dimensionless variational 
parameters. $ \alpha (t) $ is the square of the radius of the system.
In order to describe the dynamics of the variational parameter $ \alpha (t) $,
the phase factor $ i \beta (t) r^2 $ is required \cite{gar}. Note that the same kind of phase factor
is also present in the Eq. (4) of the Ref. \cite{kagan} to get the dynamics of the
width of a system. We get 
the effective Lagrangian $ L_{eff}(\alpha ,\beta ) $ by substituting Eq. (\ref{wave}) into 
Eq. (\ref{lag})  and integrating the Lagrangian density over the space coordinates,
\begin{equation}\label{Lag}
L_{eff} = \frac{ N \hbar \omega_0}{4} [- \alpha \dot \beta d + \frac{d}{\alpha} 
+ \alpha d (1 + \beta^2 ) + G_d  \alpha^{(n+m-k d)/2}],
\end{equation}
where 
\begin{equation}
  G_d   =   G_0 \int  d^{d}R  d^{m}R^{\prime} 
  e^{-(R^2 + R^{{\prime}2} )}
  V(\vec R- \vec R^{\prime }) e^{-(R^{ \prime 2 } +  R^{2}(2k-1))}, 
\end{equation}

 where $ G_0 = \frac{4 g_d N^k 2^{(d+m +n)/2} a_{0}^{(n + m-k d)}}{(k+1) \hbar \omega_0 \pi^{(k+1)d/2}} $
 and $  r^2 = 2 \alpha R^2 a_{0}^2 $.

Using the Lagrange equation of motion, we get the equation of motion for the 
variational parameter $ \alpha (t) $.
The time evolution of the variational parameter $ \alpha (t) $ is
\begin{equation}\label{alpha}
\ddot \alpha = 2\alpha \beta^2 + \frac{2}{\alpha } - 2 \alpha -
 \frac{(n+m-k d) G_d}{d} \alpha^{(n+m-k d)/2}.
\end{equation}
The $"."$ represents the derivative with respect to the dimensionless variable $ \tau = \omega_{0} t $.
The total energy of the system is
\begin{equation} 
\tilde{E} = 4E/N \hbar \omega_0 = \alpha \beta^2 d + \frac{d}{ \alpha 
} + \alpha d + G_d \alpha^{(n+ m-k d)/2}.
\end{equation}
At equilibrium, $ \beta = 0 $ and $ \frac{\partial E}{\partial \alpha}|_{\alpha = \alpha_0} = 0 $.
The equilibrium point $ \alpha_0 $ satisfies the following  relation in $d$-dimension
\begin{equation}
 \alpha_{0}^2 + \frac{(n+m-k d)G_d}{2d}\alpha_{0}^{(n+m-k d)+1} = 1. 
\end{equation}

We choose $n$, $m$, $k$ and $d$ such that they satisfy the following relation
\begin{equation}\label{main}
 n + m -k d + 2 = 0.
\end{equation}

If $n$, $m$, $k$, and $d$ satisfy the above relation, Eq. (\ref{alpha}) can be
written as 
\begin{equation}
\ddot \alpha + 4 \alpha = 2 \tilde{E}(\alpha). 
\end{equation}

Define $ I(t) = \int d^d r  | \psi ( \vec r,t )|^2 r^2 $ which is the expectation value of the square of
the radius.
The above equation can be re-written as 
\begin{equation}\label{var}
\ddot {  I } + 4 \omega_{0}^2 I = \frac{4 E(I)}{M}.
\end{equation}
From now onwards $"."$ indicates derivative with respect to the time $t$.
For monopole excitation, the average value of the collective coordinate $ r^2$ oscillates around its 
equilibrium value. We expand $I(t)$ around its equilibrium point $ I_0 $ as $ I(t) = I_0 + \delta 
I(t) $ and we get, 
\begin{equation}
\delta \ddot{I} + 4 \omega_{0}^2 \delta {I} = \frac{4 E_{0}}{M},
\end{equation}
where $ E_0 = d (\frac{1}{\alpha_0 } + \alpha_0 + \frac{G_d}{d}) $ is the equilibrium energy of the system.
This equation is identical with the classical equation (1) of Ref. \cite{pit} which was derived using 
the virial theorem.
The solution of this equation is
\begin{equation}
 I (t) = A \cos (2 \omega_0 t + \theta ) + \frac{E_0}{M \omega_{0}^2}.
\end{equation}
A and $ \theta $ are constants which can be determined from the initial conditions on $ \psi $.
Thus the square of the system radius oscillates with a frequency $ 2 \omega_0 $ in any dimension if
$ n + m - k d + 2 = 0 $.
This monopole frequency 2 $ \omega_0 $ is universal because  it is independent of 
the interatomic interaction, dimensionality of the trap, order of non-linearity and 
the total number of atoms N.
For {\em ordinary } GP equation in $ d (=m) $ dimension, $ k = 1 $. 
Hence the monopole mode frequency of an {\em ordinary } $ d $-dimensional GP equation is the universal 
frequency $ 2 \omega_0 $ if $ n = - 2 $.
Eqs. (\ref{alpha}) and (\ref{main}) explicitly shows how the scaling exponent of 
the interaction potential and the order of non-linearity gives
the universal monopole mode frequency.

If $n$, $m$, $d$ and $k$ satisfy the Eq. (\ref{main}), the equation (\ref{alpha}) 
can be re-written for time-independent
trap as well as for time-dependent trap as
\begin{equation}
\ddot{I}(t) - \frac{\dot{I}^2(t)}{2I(t)} + 2\omega^2(t) I(t) = \frac{Q_{d}}{I(t)},
\end{equation}
where $ Q_{d} = (\frac{N \hbar d}{2M})^2 (1+G_{d})$ is the positive
invariants under time evolution. This $ Q_d $ does not depend on
the frequency of the trap potential.
Let  $ X(t) = \sqrt{I(t)} $ is the condensate width.
The above equation can be written as
\begin{equation}\label{Hill}
\ddot{X}(t) +  \omega_{0}^2(t) X(t) = \frac{Q_{d}}{X^3(t)}.
\end{equation}
This is a non-linear singular Hill's equation in $ d $ dimension. 
This is an interesting result because the time evolution of the width 
can be described by the same Hill's equation in any dimension if  
 $ n + m -k d + 2 = 0 $.
The dynamics of the width is universal because the time 
evolution of the width can be described universally by the same 
non-linear singular Hill's equation in any dimension. Moreover, this Hill's
equation can be solved analytically and exactly. 
The Eq.(\ref{Hill}) can be viewed as the classical motion of a 
fictitious particle in an effective 
$ d $-dimensional potential
\begin{equation}
V_{eff} = \frac{ \omega_{0}^2(t) X^2}{2} + \frac{ Q_{d} }{2 X^2}.
\end{equation}

The general solution of Eq.(\ref{Hill}) is
$ X(t) = \sqrt{ u{^2}(t) + \frac{Q_{d}}{W^2} v^{2}(t)} $
where $ u(t) $  and $ v(t) $  are two linearly independent
solutions of the equation 
$ \ddot p + \omega^{2}(t) p = 0 $ which satisfy the initial 
conditions $ u(t_o) = X(t_0) $, $ \dot u(t_0) = X^{\prime}(t_0) $, 
$ v(t_0) = 0 $ and $ v^{\prime}(t_0) \neq 0 $. W is the  Wronskian.
This time-dependent Hill's equation can be solved explicitly only
for particular choices of $ \omega_{0}(t) $.
For time-dependent periodic trap potential, the equation is 
known as Mathieu equation which is well studied.
In most of the BEC experiment, the trap potential is time-independent.
For time-independent trap potential, a solution of this equation is
\begin{equation}\label{dyn}
X(t) = \sqrt{ \cos^2(2 \omega_0 t)+ \frac{Q_{d}}{4 \omega_{0}^2}
\sin^2(2 \omega_{0} t )}.
\end{equation}

From Eq.(\ref{main}), we get $ k d = 2$ for 1D {\em modified }
GP equation, and 2D {\em ordinary } GP equation with $ n=-2$.
This  is the condition for self-similar solution of a non-linear
GP equation discussed in by Rybin \cite{rybin,kolo}.
There is no self-similar solution in 1D {\em ordinary } GP equation
since it does not obey Eq.(\ref{main}).

We now discuss few systems which satisfy the Eq. (\ref{main}) and
shows the universal monopole mode frequency and the dynamics of the
width of the system that can be described by the Eq. (\ref{dyn}).
    
\section{ 2D TRAPPED BOSE GAS}

It has been shown by  Baganato {\em et al}. \cite{baganato} that
for an ideal 2D Bose gas  under harmonic trap a  macroscopic 
occupation of the ground state can exist at temperature 
$ T < T_c = \sqrt {N  / \zeta  (2)} \hbar \omega / k_B $.
With present technology  one can freeze the motion of the 
trapped particles  in one direction to create a quasi-2D Bose gas.  
In the frozen direction the particles execute the zero point motion. 
To achieve  this quasi-2D system, the frequency in the frozen 
direction should be much larger than  the frequency in the X-Y plane
and the mean interactions between the particles.
The wave function in the z-direction is
\begin{equation}
\psi ( z ) = C e^{- \frac{ z^2 }{2 a_{z}^2 }},
\end{equation}
where $ a_{z} = \sqrt{ \hbar/m \omega_{z}} $ is the oscillator length 
in the z-direction and $ C $ is the normalization constant.
When we integrate out  the  z-component in the 3D GP energy functional, 
then we get the effective interaction potential 
$ V_{I}= g_{2} \delta^2 (\vec r -\vec r^{\prime})$ \cite{tarun}
where $ g_{2} = 2 \sqrt{2 \pi} \hbar \omega_{z} a a_{z}  $ is the effective
coupling strength in 2D. $ a $ is the $ s $-wave scattering length in 3D.
This quasi-2D BEC state can be described
by the {\em ordinary } GP equation which is valid only when 
$ a < a_z $ \cite{pet}.
The same effective coupling constant is 
obtained in \cite{ma}. Under the scale transformation,
 $ \vec r \rightarrow \lambda \vec r $, $ V_I (\lambda \vec r ) = 
\frac{1}{\lambda^2 } V_I ( \vec r ) $.
So this interaction potential has the scaling exponent $ n =- 2 $. 
We will show that for an isotropic harmonic trapped system, the monopole 
mode frequency is $ 2 \omega_0 $. In this system,
$ m = d = 2 $ and $ k = 1 $ which satisfy the Eq. (\ref{main}).

The time-evolution of this system is
\begin{equation}\label{hill}
\ddot {X} +  \omega_{0}^2 X = \frac{ Q_{2} }{X^3 }.
\end{equation}
The same equation is obtained by using different method at
zero temperature \cite{kagan,juan}.
Here, we have identified the value of $  Q_2 $ at $ T=0 $ 
which is $ Q_{2} = \frac{N^2 \hbar^2}{M^2}(1 + \sqrt{\frac{2}{\pi}} 
N \frac{a}{a_z}) $.
The dynamics of the width can be described by the Eq. (\ref{dyn}).
At equilibrium, $ X_{0}^4 = \frac{ Q_{2} }{\omega_{0}^2 }$. When 
we linearize Eq. (\ref{hill}) around the 
equilibrium point $ X_0 $, we get,
\begin{equation}
\delta {\ddot X } + 4 \omega_{0}^2 \delta X = 0.
\end{equation}
Once again we obtain the universal oscillation frequency of the 
condensate width to be 
$ \omega = 2 \omega_0 $, corresponding to the frequency of a single
particle excitation in the condensate state.

\section{1D TRAPPED BOSE GAS}
It has been shown by Kolomeisky {\em et al}. \cite{kolo} that the GP equation 
describing BEC in 1D contains $ |\psi |^6 $ interaction term instead of familiar 
$ |\psi |^4 $ term.
The energy functional of 1D dilute interacting Bose gas at zero 
temperature is
\begin{equation}
 E = \int dx [\frac{ \hbar^2 }{2M} | \frac{ d \psi }{d x} |^2 + V(x) |\psi |^2 
 + \frac{ \hbar^2 \pi^2}{6 M } |\psi |^6].
 \end{equation}
 
In this system, $ d = m = 1 $, $ k = 2 $ which satisfy the
Eq. (\ref{main}) and $ g_1 = \frac{ \hbar^2 \pi^2}{2 M} $.
 
The time-evolution of the system is
\begin{equation}
\ddot{ X } +  \omega_{0}^2 X = \frac{Q_{1}}{X^3}.
\end{equation}

where $ Q_{1} = \frac{ N^2 \hbar^2 }{4 M^2}[ 1 + \frac{ 4 \pi N^2 }{ 3 \sqrt{3}}] $.
The same equation is obtained in \cite{rybin,kolo} by using different method.
But here we have identified the value of constant $ Q_{1} $. 
This 1D {\em modified } GP equation also exhibits an universal monopole mode frequency
$ 2 \omega_0 $ and satisfy the self-similar condition. The dynamics can be described
by the Hill's equation which is exactly solvable.  

For 1D ordinary GP equation, $ m = d = k = 1$, which does not satisfy the relation 
(\ref{main} ).
So the monopole mode frequency is affected by the interatomic strength and
number of particles and the time-evolution of the system can
not be described by non-linear singular Hill's equation.

\section{calogero model}

Only an inverse square pair potential, $ 1/r^2 $, has the scaling, $ n = -2, $
in any dimension and it satisfies the Eq. (\ref{main}) for $ k = 1 $ 
in arbitrary $ d $. A system of trapped particles interacting through pair
potential $ 1/r^2 $  is well known as the Calogero-Sutherland model \cite{calo}.
In Ref. \cite{john}, they have shown that the energy spectrum of a system of 
harmonic trapped particles interacting with $ 1/r^2 $ potential is divided into
sets of equidistant levels with separation $ 2\omega_0$ in any dimension.
In Ref.\cite{suther}, they have considered a one-dimensional system of
particles interacting by an inverse square pair potential and shown that
the monopole mode frequency is the universal frequency $2 \omega_0 $.
For $ m=k=0 $ and $ n=-2$, we get Calogero model \cite{calo} 
in arbitrary $d$ which gives the same universal monopole mode frequency 
$2\omega_0 $. In this system, the dynamics of the width can be described
universally by the Eq. (\ref{dyn}).

Apart from these two potentials, there are some other model potentials which
have the scaling exponent $n=-2$. For example, the derivative delta function
potential in 1D, the double delta function potential in 1D, and the $1/x$ delta
function potential in 1D. For all these model potentials, one can get the
universal monopole mode frequency and the dynamics of the width.

\section{ summary and conclusions }
In this work, using the time-dependent variational analysis,
we have obtained the condition, $ n + m - k d + 2 = 0 $, for 
which a class of NLSE exhibits the 
universal monopole oscillation frequency 2 $ \omega_o $.
This monopole mode frequency is universal because it does not depend on the 
strength of the interaction potential, 
dimensionality of the trap, order of non-linearity  and total number of atoms N.
We have also shown the universality of time evolution of the width in a 
time-independent trap as well as a time-dependent trap for a class of
GP equation. Interestingly,
this dynamics of the width can be described by the same Hill's equation 
(\ref{Hill}) if the Bose system satisfy the condition, $ n + m - k d + 2 = 0 $.
This Hill's equation is analytically and exactly solvable.
Using the time dependent variational analysis, we have also obtained 
the condition for self-similar solutions, $ k d = 2 $, of a 
class of non-linear GP equation.
 
As an example, we have shown that the 2D  trapped Bose gas describing by the 
{\em ordinary } GP equation, 1D BEC describing by the {\em modified } GP equation
and a system describing by the Calogero model shows the universal nature of the
monopole mode frequency and the dynamics of the width of the system. 
We have identified the values of the positive invariants quantity,
 $  Q_{2} $ and $ Q_1 $ for 2D and 1D BEC.
 
After submitting this paper on the net (cond-mat/0012188), there 
is another paper \cite{piju} appeared
 where the universality of the monopole 
mode and the dynamics of the width of a class of NLSE 
has been shown exactly and analytically which is obtained from different method.
Note that the phase factor in the variational ansatz of the wave
function (eq.(4)) is required to get the dynamics of the width \cite{gar}.
Similarly, the phase factor is also present in equation of the time 
dependent transformations in \cite{kagan,piju}. 
In Ref.\cite{piju}, using those time dependent transformations, the universality
of the monopole mode and the dynamics of the width
is obtained for a class of NLSE.
Our results obtained from the 
variational approach are identical to the exact results of \cite{piju}. 
This is because of the similarity between 
the variational ansatz for the wave function and the transformation equation
(eq.(2)) in Ref. \cite{piju}.

Our result is also valid for SO(2,1) invariant multicomponent NLSE describing BEC state.
Recently, 2D and 1D BEC has been observed in MIT \cite{gus}. Experimentally, it is also
possible to check the validity of the two-body potential in 2D BEC state
\cite{pit} and the order
of the nonlinearity in 1D BEC state by measuring the universal monopole mode frequency
and the dynamics of the width of the condensate state.

I would like to thank  Pijush K. Ghosh, M. V. N. Murthy and K. Santosh
for helpful discussions and many valuable suggestions.

\end{document}